\documentclass{elsarta}
\begin{document}

\hspace*{4 in}CUQM-108\\
\hspace*{4 in}math-ph/0504056\\
\hspace*{4 in}April 2005
\vspace*{0.4 in}
\begin{frontmatter}
\title{Perturbation theory in a framework of iteration methods}
\author[hakan]{Hakan Ciftci\thanksref{now}},
\thanks[now]{Present address: Department of Mathematics and Statistics, Concordia University,
1455 de Maisonneuve Boulevard West, Montr\'eal,
Qu\'ebec, Canada H3G 1M8.}
\ead{hciftci@gazi.edu.tr}
\author[hall]{Richard L. Hall\corauthref{cor}},
\corauth[cor]{Corresponding author.}
\ead{rhall@mathstat.concordia.ca}
\author[saad]{Nasser Saad}
\ead{nsaad@upei.ca}
\address[hakan]{Gazi Universitesi, Fen-Edebiyat Fak\"ultesi, Fizik
B\"ol\"um\"u, 06500 Teknikokullar, Ankara, Turkey.}
\address[hall]{Department of Mathematics and Statistics, Concordia University,
1455 de Maisonneuve Boulevard West, Montr\'eal,
Qu\'ebec, Canada H3G 1M8}
\address[saad]{Department of Mathematics and Statistics,
University of Prince Edward Island,
550 University Avenue, Charlottetown,
PEI, Canada C1A 4P3.}
\begin{abstract}
In a previous paper (J. Phys. A {\bf 36}, 11807 (2003)), we introduced the `asymptotic iteration method' for
solving second-order homogeneous linear differential equations. In this paper, we study perturbed problems in quantum mechanics and we use the method to find the coefficients in the perturbation series for the eigenvalues and eigenfunctions directly, without first solving the unperturbed problem.
\end{abstract}

\end{frontmatter}
\section{Introduction}\label{intro}
Boundary value problems play an important role in mathematical physics. Eigenvalue problems in quantum mechanics represent some of the most important mathematical problems in physics. Schr\"odinger's equation has exact solutions for certain special potentials but in most cases approximation methods are required. In perturbation theory the Hamiltonian $H = H_0 +\lambda V_P$ has two parts, a soluble term $H_0$ to which a perturbation $\lambda V_P$ is added. The wavefunctions and the eigenvalues of the total Hamiltonian are written as a perturbation series in the parameter $\lambda$ with coefficients calculated in terms of the exact solutions of $H_0.$ From the practical point of view, the goal of this paper is to devise an iterative method to find the coefficients in such perturbation series.

In this work, we will discuss the application \cite{chs}-\cite{brkt} of the asymptotic iteration method (AIM) not to solve the problems directly, as we have done earlier, but rather to find the perturbation coefficients.  Shortly we shall summarize the general structure of the method and show how it is applied to boundary value problems. The main purpose of the present paper is to calculate the coefficients of the perturbation expansion without using the base eigenfunctions of $H_0$ explicitly.  The differential equation of the original boundary-value problem may first have to be transformed into a certain standard form with appropriate properties, such as the presence of non-zero terms in the function and in its first derivative, or the incorporation of boundary conditions.

The paper is organized as follows. In section~2, we outline the asymptotic iteration method in section~3 we formulate the application of the method to perturbation expansions. In section~4 we apply the theory to a variety of interesting problems, and we summarize our conclusions in section~5.
\section{The asymptotic iteration method (AIM)}\label{AIM}
In this section we give the main structure of the AIM, details can be found in \cite{chs}. Given that $\lambda_0(x)$ and $s_0(x)$ are sufficiently many times continuously differentiable, the second-order differential equation
\begin{equation}\label{eq21}
y''=\lambda_0(x)y'+s_0(x)y
\end{equation}
has a general solution
\begin{equation}\label{eq22}
y(x)=exp(-\int^x\alpha dt)\bigg[C_2+C_1\int^x\exp\bigg(\int^t(\lambda_0(\tau)+2\alpha(\tau))d\tau\bigg)dt\bigg],
\end{equation}
if for some $n>0$,
\begin{equation}\label{eq23}
{s_n\over \lambda_n}={s_{n-1}\over \lambda_{n-1}}\equiv \alpha
\end{equation}
where 
\begin{equation}\label{eq24}
\lambda_n=\lambda'_{n-1}+s_{n-1}+\lambda_0\lambda_{k-1},\quad\hbox{and}\quad s_n=s'_{n-1}+s_0\lambda_{n-1}.
\end{equation}
The termination condition of the method, as given in (\ref{eq23}), can also be written as follows
\begin{equation}\label{eq25}
\delta(x)=s_{n}(x)\lambda_{n+1}(x)-s_{n+1}(x)\lambda_{n}(x) =0.
\end{equation}
This $\delta(x)$ function has a crucial role for boundary-value problems:  the coefficients $s_n$ and $\lambda_n$ now depend on the (unknown) eigenvalue $E$, which is determined by the vanishing condition (\ref{eq25}).   If, for a suitable choice of $E$, this equation is satisfied at every $x$ point, then the problem is called `exactly solvable'. All the applications in the present paper have this feature in the sense that we find exact expressions for the perturbation coefficients. The asymptotic iteration method was developed with the idea of using minimal algebraic computations to solve second order differential equations. Indeed, with the availability of symbolic mathematical software, the computations required by the asymptotic iteration (\ref{eq23}) present few difficulties, even for the higher iteration steps. AIM has proved to be useful in obtaining solutions for many boundary-value problems, including eigenvalue problems of the Schr\"odinger type \cite{chs}-\cite{chss}. 
\section{The application of AIM to perturbation problems}\label{Per_I}
\subsection{Perturbation expansions for eigenvalues}

The purpose of this section is to show that the asymptotic iteration method can be used to calculate the coefficients of a perturbation expansion for Schr\"odinger eigenproblems. We suppose that we have a potential that is the sum of two parts expressed by
\begin{equation}\label{eq311}
V(x)=V_{1}(x)+\lambda V_{2}(x),
\end{equation}
where $V_{1}(x)$ has an exact solution, $V_{2}(x)$ is the perturbation, and $\lambda$ is a perturbation expansion parameter. We suppose that the eigenvalue can be written as a perturbation series of the form 
\begin{equation}\label{eq312}
E=E^{(0)}+\lambda E^{(1)}+\lambda^{2} E^{(2)}+\lambda^{3} E^{(3)}+\dots.
\end{equation} 
Our goal is to calculate the coefficients $E^{(j)}$, where $j=0,1,2,\dots$. For the potential given in (\ref{eq311}), the correspondence Schr\"odinger equation has the form 
\begin{equation}\label{eq313}
\left(-{{d^{2}}\over{dx^{2}}}+V_{1}(x)+\lambda V_{2}(x)\right)\psi(x)=E\psi(x).
\end{equation}
After writing $\psi(x) = y_0(x)f(x)$ (and, perhaps, using additional changes of variable), we obtain a second order linear homogeneous differential equation for the factor $f(x)$ with the general form
\begin{equation}\label{eq314}
f^{\prime \prime}(x)=\lambda_{0}(x,\lambda)f^{\prime}(x)+s_{0}(x,\lambda)f(x).
\end{equation}
We assume that $E$ has the form given in (\ref{eq312}). Consequently, the $\lambda_{0}(x,\lambda)$ and $s_{0}(x,\lambda)$ functions depend on each $E^{(j)}$ too. We now apply the asymptotic iteration method to (\ref{eq314}).  After calculating the $\lambda_{n}(x,\lambda)$ and $s_{n}(x,\lambda)$ functions by using (\ref{eq24}), we can construct the $\delta(x,\lambda)$ function from (\ref{eq25}) as follows
\begin{equation}\label{eq315}
\delta(x,\lambda)=s_{n}(x,\lambda)\lambda_{n+1}(x,\lambda)-s_{n+1}(x,\lambda)\lambda_{n}(x,\lambda) =0.
\end{equation}
If we expand $\delta(x,\lambda)$ about $\lambda=0$, we get the following series
\begin{eqnarray}
\delta(x,\lambda)&=&\delta(x,0)+{{\lambda}\over{1!}}\left.{{\partial\delta(x,\lambda)}\over{\partial\lambda}}\right|_{\lambda=0}+{{\lambda^{2}}\over{2!}}\left.{{\partial^{2}\delta(x,\lambda)}\over{\partial\lambda^{2}}}\right|_{\lambda=0}
+{{\lambda^{3}}\over{3!}}\left.{{\partial^{3}\delta(x,\lambda)}\over{\partial\lambda^{3}}}\right|_{\lambda=0}+...\nonumber\\
&=&\sum_{k=0}^{\infty}{\lambda^{k}\delta^{(k)}}(x).\label{eq316}
\end{eqnarray}
According to the procedure of AIM, $\delta(x,\lambda)$ must be zero; if this is to be so for every $\lambda$ value, then every term of the series must be zero. That is to say,
\begin{equation}\label{eq317}
\delta^{(j)}(x)={{1}\over{j!}}\left.{{\partial^{j}\delta(x,\lambda)}\over{\partial \lambda^{j}}}\right|_{\lambda=0}=0,\quad j=0,1,2,\dots .
\end{equation}
It is clear that solution of the equation $\delta^{(0)}(x)=0$ gives us $E^{(0)}$ and $\delta^{(1)}(x)=0$ gives $E^{(1)},$ the first correction term to the eigenvalue. This procedure thus allows us to find the coefficients of the eigenvalue expansion. It is an attractive feature of AIM that we can also calculate the eigenfunctions, as we now show.  
\subsection{Perturbation expansions for the wave function}
When the condition given in (\ref{eq25}) is satisfied, the solution of the (\ref{eq314}) can be found using (\ref{eq22}). If we rewrite the first part of (\ref{eq22}), we have
\begin{equation}\label{eq321}
f(x)=C_{2}\exp\left(-\int\limits^{x}\alpha(t,\lambda) dt\right),
\end{equation}
where $\alpha(x,\lambda)={{s_{k}(x,\lambda)}\over{\lambda_{k}(x,\lambda)}}$, $k$ is the iteration number, and $C_{2}$ is an integration constant. Again if we expand $\alpha(x,\lambda)$ about
 $\lambda=0$, we obtain the following series
\begin{eqnarray}
\alpha(x,\lambda)&=&\alpha^{(0)}(x)+\lambda \alpha^{(1)}(x)+\lambda^{2} \alpha^{(2)}(x)+\lambda^{3} \alpha^{(3)}(x)+\dots\nonumber\\
&=&\sum_{k=0}^{\infty}{\lambda^{k}\alpha^{(k)}}(x),\label{eq322}
\end{eqnarray} 
where $\alpha^{(j)}(x)$, $j=0,1,2,...$ is given by
\begin{equation}\label{eq323}
\alpha^{(j)}(x)={{1}\over{j!}}\left({{\partial^{j}}\over{\partial\lambda^{j}}}\left({{s_{k}(x,\lambda)}\over{\lambda_{k}(x,\lambda)}}\right)\right)_{\lambda=0}={{1}\over{j!}}\left({{\partial^{j}\alpha(x,\lambda)}\over{\partial\lambda^{j}}}\right)_{\lambda=0}.
\end{equation}
The function $f(x)$ in (\ref{eq321}) becomes
\begin{equation}\label{eq324}
f(x)=C_{2}f^{(0)}(x)f^{(1)}(x)f^{(2)}(x)f^{(3)}(x)...=C_{2}\prod^{\infty}_{k=0}f^{(k)}(x),
\end{equation}
where 
\begin{equation}\label{eq325}
f^{(k)}(x)=\exp\left(-{\lambda^{k}}\int\limits^{x}(\alpha^{(k)}(t))dt\right),\quad k=0,1,2,\dots.
\end{equation}
\section{Some applications}\label{app_II}
In this section, we present some examples to demonstrate how AIM can be used to determine the perturbation coefficients.  Exact results are recovered for some well known cases. We shall discuss three examples: the quartic anharmonic oscillator, the complex cubic oscillator, and a perturbed P\"oschl-Teller potential. Next we shall obtain a perturbation expansion for the angular spheroidal eigenvalues by using the expansion for the perturbed P\"oschl-Teller potential. For perturbed harmonic oscillators we note that exactly similar treatments may be  given for the wider class of potentials of the form $V(x)=x^{2}+\lambda x^{q},$ where $q=3,4,5,\dots$. We shall first look at two specific regimes of the quartic example $q = 4.$ 
\subsection{The quartic anharmonic Oscillator:}
\subsection*{Case~1}
The potential for the quartic anharmonic oscillator may be written $V(x)=x^{2}+\lambda x^{4}.$  Several interesting studies of this problem may be found in the literature \cite{rad}-\cite{gfc}. With this potential, Schr\"odinger's equation becomes
\begin{equation}\label{eq411}
\left(-{{d^{2}}\over{dx^{2}}}+x^{2}+\lambda x^{4}\right)\psi(x)=E\psi(x),
\end{equation}
where $\lambda$ is a small parameter. After substituting $\psi(x)=\exp({-{{1}\over{2}}x^{2}})f(x)$ into (\ref{eq411}), we get
\begin{equation}\label{eq412}
f^{\prime \prime}(x)=2xf^{\prime}(x)+(1-E+\lambda x^{4})f(x),
\end{equation}
where we suppose that 
\begin{equation}\label{eq413}
E=\sum_{k=0}^{\infty}\lambda^{k}E^{(k)}.
\end{equation}
Now we can calculate the coefficients in the energy expansion. If we solve (\ref{eq317}) for $j=0$ together with (\ref{eq412}) and ({\ref{eq413}), we immediately find 
$$E^{(0)}_{n}=2n+1, \quad n=0,1,2...$$ For $j=1$  we find that 
$$E^{(1)}={{3}\over{4}},~ {{15}\over{4}},~ {{39}\over{4}},~ {{75}\over{4}}\dots.$$ Consequently, $$E^{(1)}={{3}\over{4}}(2n^{2}+2n+1).$$ 
For $j=2$, we find
$$E^{(2)}=-{{21}\over{16}},~ -{{165}\over{16}},~ -{{615}\over{16}},~ -{{1575}\over{16}},\dots$$
 that is to say  $$E^{(2)}=-{{1}\over{16}}(34n^{3}+51n^{2}+59n+21).$$ Calculating $E^{(3)}$ in a similar way, we find that $$E^{(3)}_{n}={{333}\over{64}},~ {{3915}\over{64}},~ {{20079}\over{64}},~{{66825}\over{64}}, ~{{171153}\over{64}},~ {{369063}\over{64}},\dots$$ and 
$$E^{(3)}={{3}\over{64}}(125n^{4}+250n^{3}+472n^{2}+347n+111).$$ 
Finally, we can write $E_{n}$, for $n=0,1,2,\dots$, as follows
\begin{eqnarray}
E_{n}&=&(2n+1)+{3\over{4}}(2n^{2}+2n+1)\lambda-{{1}\over{16}}(34n^{3}+51n^{2}+59n+21)\lambda^{2} \nonumber \\ 
&+&{{3}\over{64}}(125n^{4}+250n^{3}+472n^{2}+347n+111)\lambda^{3}+\dots.\label{eq415}
\end{eqnarray}
\subsection*{Case~2}
The Rayleigh-Schr\"odinger perturbation expansion is not applicable for sufficiently large value of the parameter $\lambda.$  This drawback can overcome using AIM by means of a substitution of the form $\psi(x)=\exp(-{1\over{2}}\alpha x^{2}-{1\over{4}}\beta x^{4})f(x)$. After inserting this expression for $\psi(x)$ into the (\ref{eq411}), we obtain
\begin{equation}\label{eq416}
f^{\prime \prime}(x)=2(\alpha x+\beta x^{3})f^{\prime}(x)+(\alpha-E-\beta^{2} x^{6})f(x),
\end{equation}
where $\alpha^{3}-\alpha-{3\over{2}}\lambda=0$ and $\beta={\lambda\over{2\alpha}}$. We point out that when $\lambda \geq 0$, then $\alpha \geq 1$. If we change the independent variable $x$ to $z$, where $x={{z}\over{\sqrt{\alpha}}}$, then we find (in terms of $z$)
\begin{equation}\label{eq417}
f^{\prime \prime}(z)=2(z+\eta z^{3})f^{\prime}(z)+(\epsilon-\eta^{2} z^{6})f(z),
\end{equation}
where $\eta={{\beta}\over{\alpha^{2}}}={{\lambda}\over{2\alpha^{3}}}$, and $\epsilon={{\alpha-E}\over{\alpha}}$. It is clear that $\eta$ is always less than $1$. We can write $\epsilon$ as follows
\begin{equation}\label{eq418}
\epsilon=\sum_{k=0}^{\infty}\eta^{k}\epsilon^{(k)}
\end{equation}
We now apply our perturbation procedure to this problem. If we solve (\ref{eq317}) for $j=0$, together with (\ref{eq417}) and (\ref{eq418}), we find $\epsilon^{(0)}=-2n$. For $j=1$  we find that $\epsilon^{(1)}=-3n^{2}$. For $j=2$, $\epsilon^{(2)}={{1}\over{8}}(68n^{3}+30n^{2}+64n+15)$. Finally, we can write $E_{n}$ by using $E_{n}=(1-\epsilon_{n})\alpha$ and $\eta={{\lambda}\over{2\alpha^{3}}}$  as follows
\begin{equation}\label{eq419}
E_{n}=(2n+1)\alpha +{{3n^{2}}\over{2\alpha^{2}}}\lambda-{{68n^{3}+30n^{2}+64n+15}\over{32\alpha^{5}}}\lambda^{2}+\dots.
\end{equation} 
It is clear that, (\ref{eq419}) is more strongly convergent than (\ref{eq415}) because (\ref{eq419}) includes the $\alpha$ term which is always bigger than $1$, if $\lambda > 0$.
\subsection{The complex cubic anharmonic oscillator}
In this section, we will derive a perturbation expansion for the complex cubic anharmonic oscillator. This example of a PT-symmetric potential with real eigenvalues has had much attention in the literature \cite{edt}-\cite{mga}. Specifically, we consider Schr\"odinger's equation for the potential
\begin{equation}\label{eq421}
V(x)=x^2+i\lambda x^3,\end{equation}
namely
\begin{equation}\label{eq422}
\left(-{{d^2}\over{dx^2}}+x^2+i\lambda x^3\right)\psi(x)=E\psi(x).
\end{equation}
The eigenvalues $E$ are known to be real and positive.  If we write the wave function $\psi(x)=\exp(-{{1}\over{2}}x^2)f(x)$ and substitute this form into (\ref{eq422}), we find
\begin{equation}\label{eq423}
f^{\prime\prime}=2xf^{\prime}+(1-E+i\lambda x^3)f.
\end{equation}
For small $\lambda$, the eigenvalue $E$ can be written as the perturbation series
\begin{equation}\label{eq424}
E=\sum_{k=0}^{\infty}\lambda^{k}E^{(k)}.
\end{equation}
If we solve the (\ref{eq317}) for $j=0,$ together with (\ref{eq424}) and (\ref{eq423}), we immediately find $E^{(0)}_{n}=2n+1.$ For $j=1$  we find that $E^{(1)}_{n}=0$. For $j=2$, we find $E^{(2)}_{n}={{30n^{2}+30n+11}\over{16}}.$ Calculating $E^{(3)}_{n}$ in a similar way, we find that $E^{(3)}_{n}=0$. Meanwhile for $E^{(4)}_{n}$, we must solve the equation for $j=4$ and we find $$E^{(4)}_{n}={{15}\over{256}}\left({{94n^{3}+141n^{2}+109n+31}}\right)..$$
Finally, we can write $E_{n}$ as follows
\begin{eqnarray}
E_{n}&=&(2n+1)\ +\ \left({{30n^{2}+30n+11}\over{16}}\right)\lambda^{2}\nonumber\\
&+&{{15}\over{256}}\left({{94n^{3}+141n^{2}+109n+31}}\right)
\lambda^{4} + \dots.\label{eq425}
\end{eqnarray}
This expansion is in agreement with \cite{bmw}-\cite{bmzd} and same as the result of the standard Rayleigh-Schr\"odinger perturbation expansion.
\subsection{Perturbed P\"oschl-Teller potentials}
Consider Schr\"odinger's equation 
\begin{equation}\label{eq431}
-\psi''(x)+V(x)\psi(x)=E\psi(x),\quad V(x)={{a(a+1)}\over{\sin^{2}{x}}}+\lambda \cos^{2}{x},
\end{equation}
where $ 0<x<\pi$.
After substituting $\psi(x)=\sin^{a+1}(x)f(x)$ and making the convenient change of variable $y=\cos{x}$, we find the following second order differential equation for the function $f(y)$
\begin{equation}\label{eq432}
f^{\prime\prime}=(2a+3){{y}\over{1-y^{2}}}f^{\prime}+\left({{(a+1)^2-E+\lambda y^{2}}\over{1-y^{2}}}\right)f.
\end{equation}
We assume that the eigenvalue $E$ has the expansion
\begin{equation}\label{eq433}
E=\sum_{k=0}^{\infty}\lambda^{k}E^{(k)}.
\end{equation}
If we consider $\delta^{(0)}(x)=0$ for this problem we find $E^{(0)}_{n}=(a+n)^{2}$, where $n=1,2\dots$. If we solve the equation $\delta^{(1)}(x)=0$, we obtain
$$E^{(1)}_{n}={{(2n-1)a+(n-1)(n+1)}\over{2(a+n-1)(a+n+1)}}.$$
To obtain the second correction to the eigenvalue, we have to solve
 the equation $\delta^{(2)}(x)=0$, and from this we find 
$$E^{(2)}_{n}=-{{P_{n}(a)}\over{16(a+n-2)(a+n-1)^{3}(a+n+1)^{3}(a+n+2)}},$$
where
\begin{equation}\label{eq434}
P_{n}(a)=u_{5}a^{5}+u_{4}a^4+u_{3}a^{3}+u_{2}a^{2}+u_{1}a+u_{0},
\end{equation}
and for $n = 1,2,3,\dots$
\begin{eqnarray*}
u_{0}&=&{1\over{2}}(n-2)(n-1)^{2}(n+1)^{2}(n+2)\\
u_{1}&=&-3(n-1)(n+1)(n^3-n^2-3n+1)\\
u_{2}&=&-{1\over{2}}(9n^{4}-24n^{3}-19n^{2}+24n+10)\\
u_{3}&=&2n^{3}+13n^{2}-5n-13\\
u_{4}&=&8n^{2}+2n-9\\
u_{5}&=&2(2n-1).
\end{eqnarray*}
Thus we obtain the perturbation series for the eigenvalues as follows
\begin{eqnarray}
E_{n}&=&(a+n)^2+{{(2n-1)a+(n-1)(n+1)}\over{2(a+n-1)(a+n+1)}}\lambda\nonumber\\
&-&{{P_{n}(a)}\over{16(a+n-2)(a+n-1)^{3}(a+n+1)^{3}(a+n+2)}}\lambda^{2}+\dots.\label{eq435}
\end{eqnarray}
\subsection{The angular spheroidal eigenvalues}
From (\ref{eq435}), we can get an explicit formula for the angular spheroidal eigenvalues. This important problem has a long history \cite{sld}-\cite{bog}. The angular spheroidal wave equation can be written in following form \cite{brkt}
\begin{equation}\label{eq441}
f^{\prime\prime}=(2m+2){{y}\over{1-y^{2}}}f^{\prime}+\left({{c^{2}y^{2}-\epsilon}\over{1-y^{2}}}\right)f,
\end{equation}
with $-1\leq y \leq 1$. Here, $\epsilon=(\lambda_{l}^{m}(c))^{2}-m(m+1)$, $l$ is the angular momentum quantum number, and $m$ is the eigenvalue of the operator $L_{z}$. We note that for $c=0$, $(\lambda_{l}^{m}(c))^{2}=l(l+1)$. When we compare (\ref{eq441}) with (\ref{eq432}), we find that $a=m-{1\over{2}}$, $\lambda=c^{2},$ and $\epsilon=E_{n}-(a+1)^2$. After substituting $n=l-m+1$ and $a=m-{1\over{2}}$ into the (\ref{eq435}), we obtain the perturbation expansion for $\lambda_{l}^{m}(c)$ shown below: this agrees with the results of \cite{gui} for the special case $m=0$. 
\begin{eqnarray}
(\lambda^{m}_{l}(c))^{2}&=&l(l+1)+{{(2l-2m+1)(2m-1)+2(l-m)(l-m+2)}\over{(2l-1)(2l+3)}}c^{2} \nonumber\\
&-&{{P_{l}(m)}\over{2(2l-3)(2l-1)^{3}(2l+3)^{3}(2l+5)}}c^{4}+\dots\label{eq442}
\end{eqnarray}
where $$P_{l}(m)=u_{5}(2m-1)^{5}+2u_{4}(2m-1)^4+4u_{3}(2m-1)^{3}+8u_{2}(2m-1)^{2}+16u_{1}(2m-1)+32u_{0},$$ 
and
\begin{eqnarray*}
u_{0}&=&{1\over{2}}(t-1)t^{2}(t+2)^{2}(t+3)\\
u_{1}&=&-3t(t+2)(t^3+2t^2-2t-2)\\
u_{2}&=&-{1\over{2}}t(9t^{3}+12t^{2}-37t-50)\\
u_{3}&=&2t^{3}+19t^{2}+27t-3\\
u_{4}&=&8t^{2}+18t+1\\
u_{5}&=&2(2t+1)
\end{eqnarray*}
where $t=l-m$.
\section{Conclusion}
The asymptotic iteration method allows one to iterate a second-order linear differential equation so that at each iteration an equation of the same general type is recovered. This iteration scheme applies to boundary value problems provided the boundary conditions are accommodated by a factored form for the wave function that is asymptotically correct. An exact eigenvalue is obtained when the energy $E$ can be chosen so that the iteration converges in a finite number of steps. In more general cases an approximation is obtained by forced numerical convergence.  In this paper we have applied the method to some perturbation problems and have been able to find the coefficients in the perturbation series directly, without first solving the unperturbed problem. In many cases, such as those studied here, we have been able to derive these coefficients exactly.

\section{Acknowledgments}
\medskip
\noindent Partial financial support of this work under Grant Nos. GP3438 and GP249507 from the 
Natural Sciences and Engineering Research Council of Canada is gratefully 
acknowledged by two of us (respectively [RLH] and [NS]).

\end{document}